\documentclass[structabstract]{aa}  
\usepackage{graphicx}
\usepackage{txfonts}
\usepackage{natbib}
\begin{document}
%
   \title{Time evolution and rotation of starspots on CoRoT-2 from the modelling of transit photometry \thanks{CoRoT is a space project operated by the French Space Agency, CNES, with participation of the Science Programme of ESA, ESTEC/RSSD, Austria, Belgium, Brazil, Germany, and Spain.} 
}

 \titlerunning{Evolution and  rotation of starspots on CoRoT-2}
    \authorrunning{A.~Silva-Valio and A.~F.~Lanza}

   \subtitle{}

   \author{Adriana Silva-Valio
          \inst{1}
          \and
          A. F. Lanza\inst{2} 
          }

   \institute{CRAAM, Mackenzie University, Rua da Consola\c c\~ao, 896, 01302-907, S\~ao Paulo, Brazil\\
              \email{avalio@craam.mackenzie.br}
         \and INAF-Osservatorio Astrofisico di Catania, Via S. Sofia, 78, 95123 Catania, Italy            
                     }

   \date{Received ; accepted }

 
  \abstract
   {CoRoT-2, the second planet-hosting star discovered by the CoRoT satellite, is a  young and active star. A total of 77 transits were observed for this system over a period of 135 days. }
   {Small modulations detected in the optical light curve of the planetary transits are used to study the position, size, intensity, and temporal evolution of the photospheric spots on the surface of the star that are occulted by the planetary disk. }
   {We apply a spot model to these variations and create a spot map of the stellar surface of CoRoT-2 within the transit band for every transit. From these maps, we estimate the stellar rotation period and obtain the longitudes of the spots  in a reference frame rotating with the star. Moreover, the spots temporal evolution is determined. This model achieves a spatial resolution of $2^\circ$.
 }
   {Mapping of 392 spots vs. longitude indicates the presence of a  region free of  spots, close to the equator, reminiscent of the coronal holes observed on the Sun during periods of maximum activity. With this interpretation,  the stellar rotation period within the transit latitudes of $-14 \fdg 6 \pm 10^{\circ}$ is obtained from the auto-correlation function of the time integrated spot flux deficit, yielding a rotation period of 4.48 days. With this period, the temporal evolution of the spot surface coverage in  individual $20^{\circ}$ longitude bins has periodicities ranging from 9 to 53~days with an average value  of $31 \pm 15$~days. On the other hand, the longitude integrated spot flux, that is independent of the stellar rotation period, oscillates with a periodicity of $17.7$ days, which false-alarm probability is $\sim 3$\%.
  }
   {The rotation period of 4.48 days obtained here is shorter than the 4.54 days as derived from the out-of-transit light modulation. Since the transit data  samples a region close to the stellar equator, while the period determined from out-of-transit data reflects the average rotation of the star, this is taken as an indication of a latitudinal differential rotation of about 3\% or 0.042 rad/d.  }


   \maketitle
%

\section{Introduction}

Rotational modulations of the light curves of stars have been interpreted as due to the presence of starspots on the stellar surface \citep[e.g., ][]{kron47}. Despite being first detected on red dwarfs, nowadays spot activity has been observed on all main-sequence late-type stars and is attributed to photospheric magnetic fields, by  analogy with sunspots. By modelling the rotational modulation of the stellar flux, it is possible to infer the brightness and relative sizes of starspots. Moreover, long-term data series provide information on stellar cycles, similar to the 11-year solar cycle, spot longitude distributions, and lifetimes \citep[e.g., ][]{berdyugina05,strassmeier09}.

The spot modelling of the disk-integrated light modulation, however, only allows for a few large spots on the stellar surface. This changed with the discovery of the first planetary transits of HD 209458 \citep{charbonneau00, henry00}. When the planet eclipses the star, it may occult solar-like spots on the stellar surface, producing detectable variations in the light curve during the transits. \citet{silva03} developed a model to infer the properties of starspots from transit  photometry. This model was later applied to the light-curve of CoRoT-2 \citep{silva-valio10}, the second planet-hosting star discovered by the CoRoT satellite. Here we give continuity to our analysis of the spots on CoRoT-2, following the spots in a reference frame that rotates with the star.

This same star had its long-term photometric out-of-transit data analysed by \citet{lanza09}. The authors concluded that the star had two preferred longitudes  on opposite hemispheres, spot lifetimes of the order of a month, and derived a lower limit for the amplitude of the surface  differential rotation of only 0.7\% from the migration of the active longitudes.

Furthermore,  both the rotation modulation and transits of the CoRoT-2 light-curve were modelled by \cite{huber10} with a low angular resolution, reaching $15^{\circ}$ during transits. Their modelling yields a rotation period within the transit band of 4.55 days,  while a longer period is  suggested for the non-eclipsed latitudes giving a latitudinal differential rotation  $\Delta \Omega \ga 0.1$ rad/d. The authors also find two preferred longitudes  on opposite hemispheres. Here we focus on the modelling of the photometric modulations  produced by starspots during transits, aiming at a significantly  higher longitude resolution than \citet{huber10}, thanks to the method introduced by \citet{silva03}. This approach allows a better characterization of the properties of the individual starspots along the occulted band, detecting spots as small as $\sim 0.2$ planetary radii, i.e., $2^{\circ}$ of longitude on the stellar surface \citep[][see also Sect.~\ref{result} below]{silva-valio10} and  using them  as tracers of stellar rotation. 

By tracking the motion of sunspots across the disc of the Sun, \cite{carr1863} noticed that their sidereal angular velocity  could be approximated by:

\begin{equation}
\Omega = A - B \cos^2 \theta, 
\label{eq:omega}
\end{equation}

\noindent where $A = 14 \fdg 55$ day$^{-1}$, $B = 2 \fdg 87$ day$^{-1}$, and $\theta$ is the co-latitude \citep{thomas08}. According to this equation, the solar equator rotation period is 24.7 days, that is, faster than that of the poles of 30.8 days. Evidence of differential rotation in other late-type stars has also been found \citep[e.g., ][]{barnesetal05}. 

The differential rotation of distinct solar activity features is not the same, but depends on the features themselves.
For instance, solar coronal holes, and their photospheric counterparts represented by magnetic unipolar regions, usually exhibit a remarkably rigid rotation in comparison to active regions and sunspot groups.  
We find something reminiscent of this behaviour in the rotation of the activity features in the photosphere of CoRoT-2 because the active longitudes trace a pattern rotating much more rigidly than the individual spots (see Section~\ref{spotlong}).

The next section provides an overview of CoRoT-2 observations, whereas the following section presents the method used. Previous results are stated briefly in Section~\ref{result}. Sections~\ref{spotlong} and \ref{spotime} describe the spots longitude and the spot temporal evolution, respectively, once the stellar rotation period is calculated. The next section explains how the stellar differential rotation is obtained, and finally Section~\ref{discussion} presents the discussion and conclusions.


\section{Observations of CoRoT-2}

The Convection Rotation and planetary Transits (CoRoT) space telescope, launched in December 2006, has discovered 15 planets so far using the method of  transits (see, e.g., http://exoplanet.eu).
The second planet discovered by CoRoT orbits a young and active star similar to the Sun, now named CoRoT-2, of spectral type G7V with mass and radius of 0.97~$M_{\odot}$ and 0.902~$R_{\odot}$, respectively. 

CoRoT-2  was continuously monitored for about 135 days, during which 77 transits of its planet were observed with high temporal resolution (32~s). The first two transits were discarded because of the lower temporal resolution. The data were reduced as explained in \cite{alonso08} with a resulting rms of  the points of the  white light curve  of  0.0006 in units of the relative flux of the out-of-transit data.
The CoRoT white passband ranges from 300 to 1100 nm and its transmission profile can be found in \citet{auvergneetal09}, together with a detailed description of the instrument,  its operation, and  the data processing pipeline.

The basic assumptions and the parameters of the planet and its orbit used for the present modelling  were taken from previous observations \citep{alonso08}, and are listed below:

\begin{itemize}
\item Circular planetary orbit of radius 6.7 $R_s$, where $R_s$ is the stellar radius, and  period of 1.743 days;
indeed, \citet{gillonetal10} find that the orbit is slightly eccentric with $e \cos \omega = -0.0029 \pm 0.0006$, where $e$ is the eccentricity and $\omega$ the argument of periastron, but this small eccentricity does not significantly affect the results derived from a simple model with a circular orbit;
\item Orbital inclination angle of $87 \fdg 84 $, implying a projection of the central transit chord at a stellar latitude of $ -14 \fdg 627$ (the Southern latitude is arbitrarily chosen);
\item The stellar equator and the planetary orbit are coplanar; this is supported by the measurement of the Rossiter-McLaughlin effect that yields a misalignment of the projected spin and orbital angular momenta of only $7^{\circ} \pm 5^{\circ}$ \citep{bouchy08};
\item Planet radius of $R_p =$ 0.172 $R_s$; this value was estimated from the deepest transit, that is, the one with  the minimum number of spots \citep{silva-valio10}.
\end{itemize}


\section{The Method} \label{method}

\begin{figure}
   \centering
  \includegraphics[width=9cm]{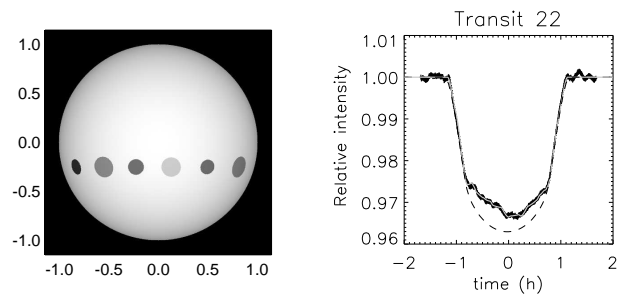}
      \caption{Example of the fit to transit 22. Left: Star with modelled nine spots. Right: observed light curve of the transit (black) with the resulting light curve from the model (thick grey line).}
         \label{star}
   \end{figure}

The model used here simulates a star with quadratic limb darkening as a 2-D image such as the one shown in the left panel of  Figure~\ref{star}. A detailed description of the model is given in \cite{silva03}. Applications of such a  model are described in \cite{silva-valio08} and \cite{silva-valio10}, the latter specifically for CoRoT-2. The model simulates the passage of the planet, a completely dark disk, in front of the spotted star. The light curve is computed as the sum of the fluxes coming from all the pixels in the frame, for a sequence of planet positions along its orbit at intervals of 2 minutes.

Unless otherwise stated, the planet and star parameters are taken from  \cite{alonso08}. Moreover, the model for the spots assumes that:

\begin{itemize}
\item The spots are circular and  described by three parameters: their radius (in units of planet radius, $R_p$), intensity (in units of stellar photosphere maximum intensity at disc centre, $I_c$), and longitude;
\item The spot latitude is the same as that of the central transit chord on the disk of the star, that is, $-14 \fdg 6$;
\item The spot longitude is the stellar topocentric longitude, i.e., with zero longitude on the  meridian at the centre of the disk of the star at mid-transit; an accuracy up to $\sim 1^{\circ}$ can be reached in the case of CoRoT-2 \citep[cf., e.g., ][]{wolteretal09};
\item The spots are constrained to longitudes $\pm 70^\circ$,  to exclude  the distortions on the ingress and egress  branches of the transit profile that cannot be measured as accurately as those falling in the central part of the transit because of the steepness of the light variation \citep[see][]{silva-valio10};
\item The effect of foreshortening when a spot is close to the limb is taken into account.
\end{itemize}

This model was  applied to all transit light curves and the three parameters (radius, intensity, and longitude) for each spot were sought simultaneously by minimizing the $\chi^{2}$ between the model light curve and the data. All fits were performed using the AMOEBA routine \citep{press92}. Initial guesses of the parameters were obtained through the genetic algorithm PIKAIA \citep{charbonneau95}. 

The minimum number of spots per transit, $\bar{M}$, needed to fit the light curve within the data uncertainty, was estimated as follows. 
A model with $M$ spots has $3 M$ free parameters, because each spot corresponds to 3 free parameters (longitude, radius, and intensity). Therefore, the number of degrees of freedom of the model is $s=N-3 M$, where $N$ is the number of data points per transit (usually $N=217$).
Now, the problem is to find the minimum number of spots  that yields an adequate best fit for each transit according to the criterion for hypothesis testing given by, e.g., \citet{lampton76}. We compute the minimum value of the $\chi^{2}$ for each model with a different $M$ starting from the case with a single spot ($M=1$). A model with $M$ spots is rejected at a significance level $\gamma$ if its minimum chi square $\chi^{2}(M)$ exceeds $\chi^{2}_{s}(\gamma)$, i.e., the $\gamma$-point of the $\chi^{2}_{s}$ distribution defined by:
\begin{equation}
\gamma \equiv \int_{\chi^{2}_{s}(\gamma)}^{\infty} f(\chi^{2}) d \chi^{2},
\label{chi_eq}
\end{equation} 
where $f$ is the density  of the probability distribution of the $\chi^{2}$ with $s$ degrees of freedom. By increasing $M$, the minimum  $\chi^{2}(M)$ decreases until for a certain value of $M$, say $\bar{M}$, we have $\chi^{2}(\bar{M}) \leq \chi^{2}_{s}(\gamma)$. The minimum number of spots giving an acceptable model is therefore $\bar{M}$. Following \citet{lampton76}, we assume a significance level $\gamma =0.1$ and compute the $\chi^{2}$ cumulative distribution function in Eq.~(\ref{chi_eq}) by means of the IDL function CHISQR\_PDF.

\begin{figure}
   \centering
  \includegraphics[width=8cm]{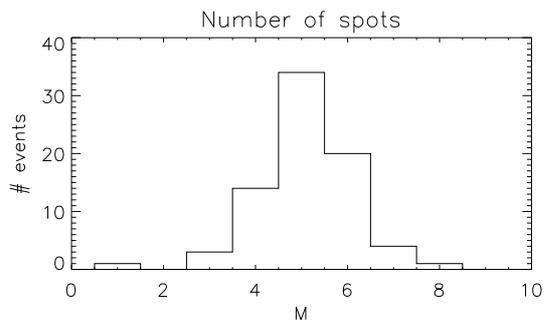}
      \caption{Histogram of the minimum number of spots per transit as derived by the method described in Sect.~\ref{method}.}
         \label{nspot}
   \end{figure}

Figure~\ref{nspot} shows the distribution of the best number of spots per transit, that is, the minimum number of spots that makes the fit statistically acceptable. A total of 392 spots were modelled during the whole period of observations (77 transits). The average number of spots per transit was $5 \pm 1$. 

In principle, our modelling approach can suffer from the effects of possible systematic errors, but we believe that their impact is  negligible because: a) the model has enough degrees of freedom to account for a highly complex flux modulation during transits by increasing the number of small spots; b) the effect of bright faculae is not detected in our transit profiles because we never observe dips below the reference profile as derived from the deepest transit; c) systematic errors  arising from  limb-darkening parameters  or the adopted relative radius of the planet are minimized  by considering only the  central part of each transit profile ($\pm 70^\circ$). In any case, our choice of a confidence level  $\gamma =0.1$ excludes too small $\chi^{2}$ values that may imply an over-fitting of the data or enhance the impact of any residual systematic error. 


\section{Spots physical parameters} \label{result}

The location in time of a ``bump"  along the transit light curve provides unambiguous information on the longitude of the corresponding spot. When the spot latitude is fixed a priori, the radius and the intensity of the spot may show some degree of degeneracy, but in the case of a highly accurate light curve, such as for CoRoT-2, they are independently constrained by the duration and the height of the corresponding bump, respectively. This is confirmed by the analysis of \citet{wolteretal09} who considered also spots centred at latitudes different from that of the central transit chord, thus studying the effect of this additional degree of freedom in the model.  

A quantity proportional to the blockage of stellar emission due to the spot, called relative flux deficit, can be defined as $D = (1 - f_i)\ S^2$, where $f_i$ and $S$ are the spot intensity (in units of $I_c$) and radius (in units of $R_p$), respectively. The deficit is proportional to $1 - f_i$ because darker spots are the predominant ones. In fact, a value of $f_i = 1$ means no spot at all, since there is no contrast with respect to the photosphere. The  advantage of using the flux deficit, instead of the values of $f_i$ and $S$, is that it is virtually independent of the degeneracy between these parameters. 

\begin{figure}
   \centering
  \includegraphics[width=9cm]{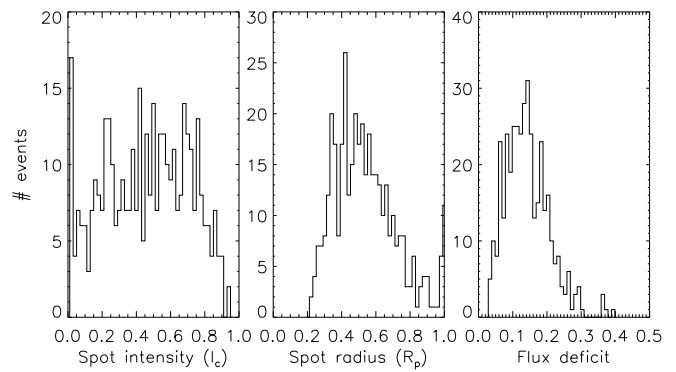}
      \caption{Histograms of the spot parameters obtained from the fits to the 77 transits: spot intensity (in units of $I_c$) (left), spot radius (in units of $R_p$) (middle), and flux deficit due to spot (right). }
         \label{res}
   \end{figure}

Figure~\ref{res} shows the results of the spot intensity, radius, and flux deficit for the spots detected during all transits, a total of 392 spots. These are different from the results reported in \cite{silva-valio10}, where a maximum of 9 spots was assumed. Here, the number of spots fitted was smaller, about half than previously used, owing to the criterion applied to find the optimal number of spots (cf. Sect.~\ref{method}). The new model gives  mean values of the spot intensity (Figure~\ref{res}, left panel) and the radius (middle panel) of $(0.45 \pm 0.25)~ I_c$, and $(0.53 \pm 0.18)~ R_p$, respectively. The corresponding distribution of the relative flux deficit is shown in the right panel of Figure~\ref{res} and has an average value of $0.14 \pm 0.06$. Because fewer spots per transit were considered in the new fits, these average values are larger than those reported in \cite{silva-valio10}.

Simulations for a single central spot with different radius and intensity show that only the spots with  a flux deficit greater than approximately 0.02 produce a signal in the light curve above the data noise level of 0.0006. Therefore, only the spots with flux deficit above 0.02 are considered in the following analysis and have their best-fit parameters plotted in Figure~\ref{res}. Note that this selection criterion implies a threshold of 0.20 $R_p$ for the spot radius and a maximum value of 0.92 $I_c$ for the spot intensity.

\begin{figure}
   \centering
  \includegraphics[width=9cm]{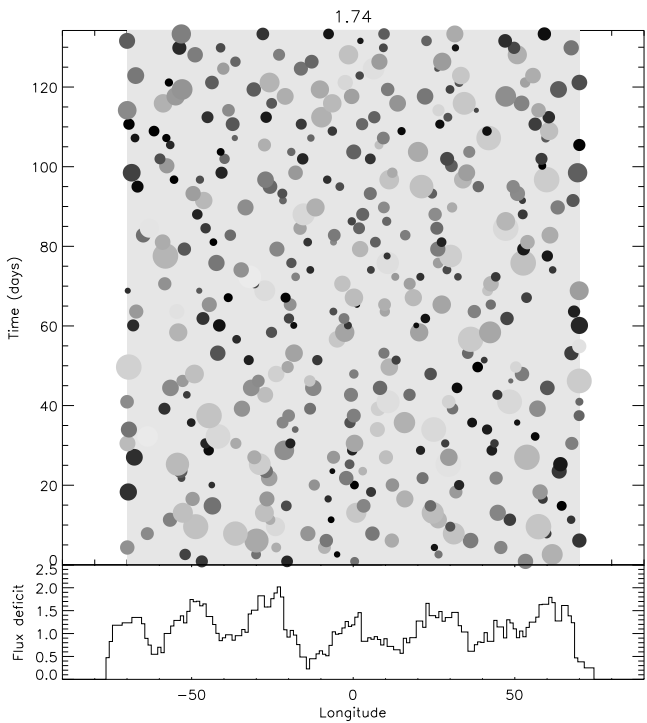}
      \caption{Top: Diagram of the spots longitude, size, and intensity for each transit. Darker spots are pictured in darker shades of grey. Bottom: Spots flux deficit integrated in time as a function of spot longitude.}
         \label{spots}
\end{figure}

The longitudes of the spots obtained from the best fit to each transit are shown in Figure~\ref{spots} where the size of the symbols represents the relative diameter of the spots, and their colour their intensity. Black spots are those with the highest contrast. Only spots within $-70^{\circ}$ and $+70^{\circ}$ longitude and flux deficit greater than 0.02 are plotted. The bottom panel shows a histogram of the flux deficit vs. longitudes for these spots.

\section{Spot longitudes and stellar rotation} \label{spotlong}

The longitude  of a spot obtained from the fit is the topocentric longitude on the star measured in an inertial  reference frame. To convert this value into a longitude measured in a reference frame that rotates with the star, that is, fixed to the stellar surface, one needs the rotation period of the star. \cite{alonso08} obtained a value of 4.54 days for the mean stellar rotation period, whereas \cite{lanza09} found $4.522 \pm 0.024$ days from the motion of the active longitudes.

The rotational longitude, $\beta_{\rm rot}$, is determined from the topocentric longitude, $\beta_{\rm topo}$, as:

\begin{equation}
\beta_{\rm rot} = \beta_{\rm topo} - 360^\circ\ {n\ P_{\rm orb} \over P_{\rm star}}, 
\label{betarot}
\end{equation}

\noindent where $P_{\rm orb} = 1.743$ day is the planet orbital period, $P_{\rm star}$ the stellar rotation period, and $n$ the transit number. The rotational longitude is limited to the range  $\pm 180^\circ$ and  coincides  with the topocentric longitude at the mid-time of the first transit, i.e., when $n=0$ at HJD 2454242.7666. 

\subsection{Comparison with the out-of-transit spot modelling}

\begin{figure}
   \centering
  \includegraphics[width=10cm]{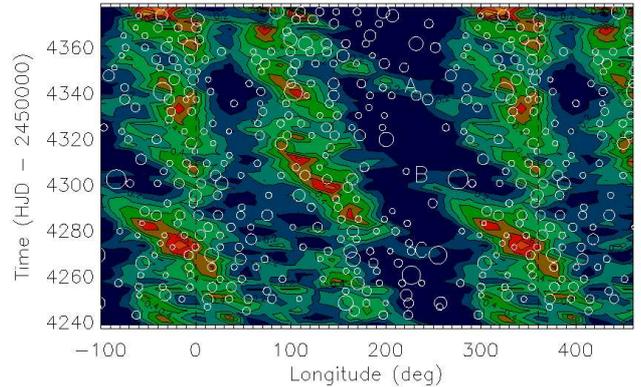}
      \caption{Spots longitude versus time for spots with flux deficit larger than 0.1 are shown as white circles overplotted on the  map of the spot filling factor obtained by \cite{lanza09} from out-of-transit data (their Fig. 4). Different colours indicate different filling factors, with yellow  indicating the maximum spot occupancy and dark blue the minimum. The bridges connecting active longitudes are labelled with A, B, and C (see the text). }
         \label{lanza}
\end{figure}

Before determining the best rotation period within the transit band, it is interesting to compare the longitudes of the spots obtained here with  the out-of-transit spot maps as derived  by \cite{lanza09}. First, the spot  topocentric longitudes shown in Figure~\ref{spots} were converted to rotational longitudes for a stellar period of 4.522 days. Fixing the first transit at HJD 2454242.7666,  the shift of the origin of the longitudes between our reference frame and that of \cite{lanza09} is $-116 \fdg 03$.
Thus, our longitudes were  shifted by $-116 \fdg 03$ to overplot both results. 
 Figure~\ref{lanza} is the same as Fig. 4 of \cite{lanza09}, with the white circles representing the spots  from the transit fits  with  a flux deficit greater  than 0.1. 
The radius of each circle is proportional to the flux deficit of the corresponding spot. The agreement between the results of the two different modelling approaches is remarkable. 

Some of the small and medium spots seem to trace small bridges between the active longitudes, even if they are not perfectly superimposed with the Maximum Entropy bridges marked with A, B, and C in Figure.~\ref{lanza}.  If  interpreted  as the effect of moving spots,  these bridges can be used to  derive  the differential  rotation {\citep[cf., e.g., ][]{frohlich09}}.  However, the transit maps suggest  that they are not single migrating spots but instead are locations where spots appear and disappear between the two active longitudes, like ``hot spots" where activity is preferentially located between the active longitudes. 
Of course, not all the spots found from transit mapping are reproduced in the out-of-transit maps because short-lived spots do not produce a significant rotational modulation signal, while  the main features found from the out-of-transit data are also found by transit mapping, including some evidence of those small bridges.
Moreover, the spatial resolution of the transit mapping is higher than that of the out-of-transit modelling which is limited to approximately $30^{\circ}-50^{\circ}$. Therefore, the spots mapped during transits show a more discrete distribution vs. longitude.

\subsection{Rotation period at the transit latitude}

\begin{figure}
   \centering
  \includegraphics[width=9cm]{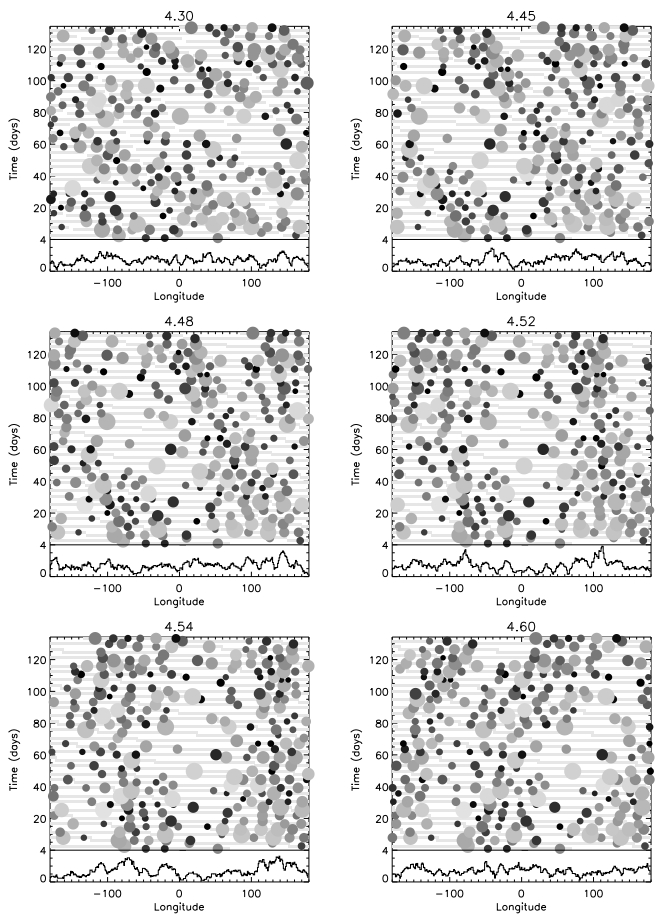}
      \caption{Spot rotational longitudes for 6 different rotation periods (reported at the top of each panel) for the spots of all transits with flux deficit larger than 0.1. The size of the circles is proportional to the spot radius whereas the colour refers to the spots flux deficit. The light grey bands in the background represent the stellar longitudes visible to the observer. Bottom: Spots flux deficit integrated over  time as a function of spot longitude.}
         \label{dif}
\end{figure}

The rotation period of the transit band needs not be equal to the star average rotation period, because the star may present differential rotation. To determine the rotation period of the stellar surface around the transit band, which covers $-14 \fdg 6 \pm 10^{\circ}$ latitude, spot maps of the stellar surface within the transit latitudes were constructed, similar to those of Figure~\ref{spots}. 
Periods ranging from  4 to 6 days were tried, with 0.01 day  intervals. 
Figure~\ref{dif} shows such maps for the periods of 4.30, 4.45, 4.48, 4.52, 4.54, and 4.60 days. These maps show the relative size and flux deficits of the spots as a function of their rotational longitude (horizontal axis) and transit time (vertical axis).

 At first glance, one sees that the spots follow diagonal stripes in these panels, giving the impression that they are wrapped around the star too tightly or too loosely. That is to say, the stellar periods used are not appropriate for the latitude of the belt occulted during transits. 
 A visual inspection of these panels shows that it is as if the spots of the initial diagonal stripes (slanted to the left) are being straightened out to an almost vertical temporal evolution (for $P_{star}$ around 4.54 days) and then the diagonal inclination changes sign (slanted to the right) for periods longer than 4.54 days. Similar to adjusting the image of an old TV set.

If the stellar rotation period at $-14 \fdg 6$ latitude is taken as 4.54 days, this would imply that CoRoT-2 has almost no differential rotation since this  practically coincides with the average period of the star as measured by \cite{lanza09}. 
The same result has been obtained by \cite{huber10} as they derived the rotation period by minimizing the drift of the active longitudes.

 Nevertheless, a different interpretation is possible
 by analogy with the rotation of solar coronal holes. Specifically, the 
blank range between 0$^{\circ}$ and 40$^\circ$ longitude, free of spots (see Figure~\ref{dif}), can be interpreted as analogous with a solar coronal hole that extends to $-14 \fdg 6$ latitude and lasts for over 120 days. From solar studies, it is known that coronal holes rotate more rigidly than sunspot groups, and with an equatorial rotation period similar to the average solar rotation period (cf. Sect.~\ref{discussion}). 
If we consider this to be the case here, we can trace the rotation of individual spots as they move across the active longitudes and the blank range.

\begin{figure}
   \centering
  \includegraphics[width=9cm]{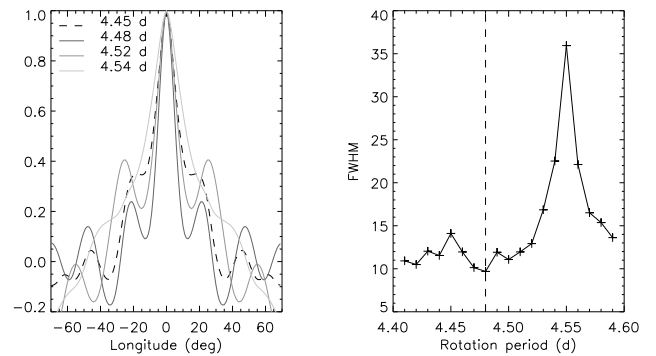}
      \caption{Left: Auto-correlation function of the spot flux deficit vs. longitude for four different stellar rotation periods as indicated by the different curves. Right: Width of the main peak at half the maximum value (FWHM) of the auto-correlation function versus stellar rotation period. } 
         \label{acor}
\end{figure}

Therefore, concentrating on individual spots, by visual inspection it may be possible to identify several lines of beads. Because of the large quantity of spots, we used only those with a flux deficit greater than 0.1 for this purpose. 
The possibility of tracing the longitude of individual spots vs. time is limited by the fact that a homogeneous longitudinal coverage of the occulted band requires at least five transits \citep[cf., e.g., Fig.~2 of ][]{huberetal09}, i.e., $\approx 9$ days during which spots generally evolve remarkably both in radius and intensity. Therefore, a quantitative criterion to find the best rotation period is to look at the auto-correlation function of the  time integrated flux deficit as a function of longitude, for the most promising periods. They correspond to the maps with vertical strings of spots, i.e., showing the minimum drift of the individual spots vs. time.

Figure~\ref{acor} (left panel) shows the auto-correlation functions for periods of 4.45, 4.48, 4.54 and 4.55~days. 
The right panel of Figure~\ref{acor} displays the full width at half maximum  (FWHM) of the main peak of the auto-correlation function versus the rotation period from 4.4 to 4.6 days. As can be seen from the figure, the period that yields the narrowest auto-correlation function is that corresponding to 4.48~d, indicated by the vertical dashed line.

\begin{figure}
   \centering
  \includegraphics[width=8cm]{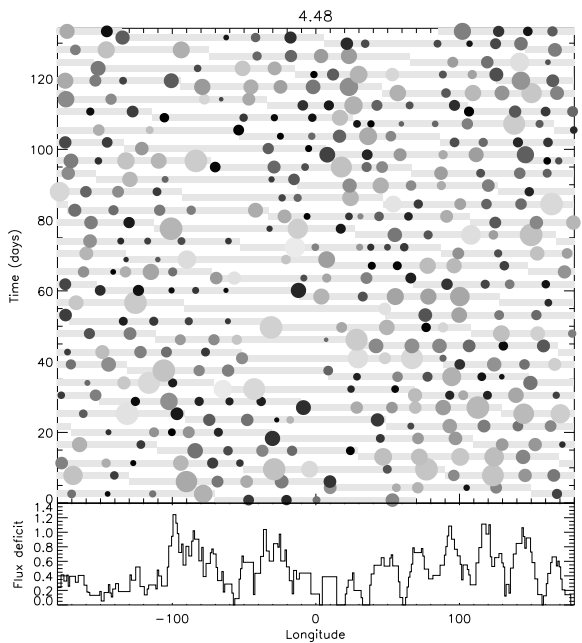}
      \caption{Surface map (top) for spots with flux deficit larger than 0.02 and time integrated flux deficit of the spots versus longitude (bottom) for a stellar  rotation period of 4.48 days. The light grey bands in the background of the top panels represent the stellar longitudes visible to the observer.  }
         \label{long}
\end{figure}

Therefore, hereafter the adopted stellar rotation period in the latitude band occulted by the planet is 4.48 days. A map of the surface of the star for each transit as a function of the rotational longitude for such a rotation period is plotted in Figure~\ref{long} (upper panel).

\section{Spot temporal evolution} \label{spotime}
\label{time_evolution}

\begin{figure}
   \centering
  \includegraphics[width=8cm]{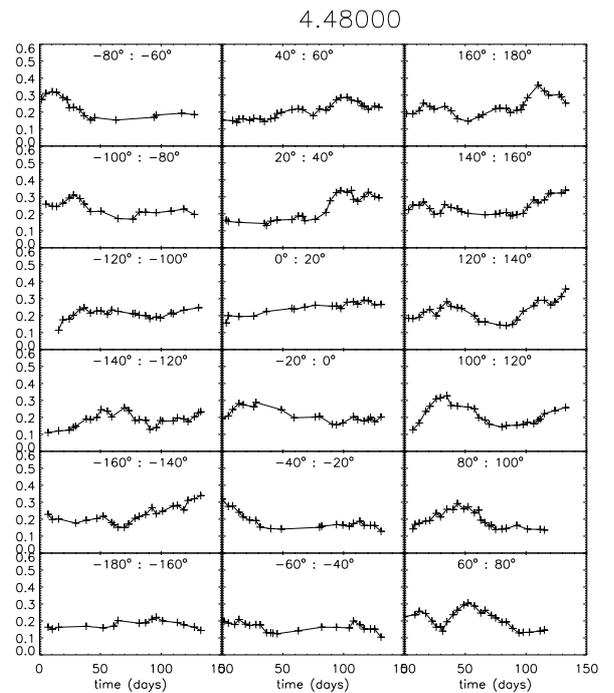}
      \caption{Evolution of the smoothed flux deficit for the 18 longitude bins.}
         \label{lifetime}
   \end{figure}

After the stellar rotation period and thereafter the rotational longitudes of the spots have been determined, it is possible to estimate the spot  temporal evolution. Here the rotational longitudes are calculated by considering the star rotation period within the occulted latitude band to be 4.48 days. First, the surface of the star is divided into 20$^\circ$ longitude bins. Then the total flux deficit  within each bin is calculated as a function of transit time. The resulting flux deficit in each bin is plotted in Figure~\ref{lifetime}, where a running mean of every five data points has been used to warrant a homogeneous sampling of all the longitudes.

The temporal evolution of the spot surface coverage within each longitude bin is estimated by computing the Lomb-Scargle periodogram of the time profiles of Figure~\ref{lifetime} with a maximum false-alarm probability of 0.01 (99\% significance). The periods obtained from the time series analysis of the data range between 9 and 53 days, with a mean value of $31 \pm 15$ days.

\begin{figure}
   \centering
  \includegraphics[width=8cm]{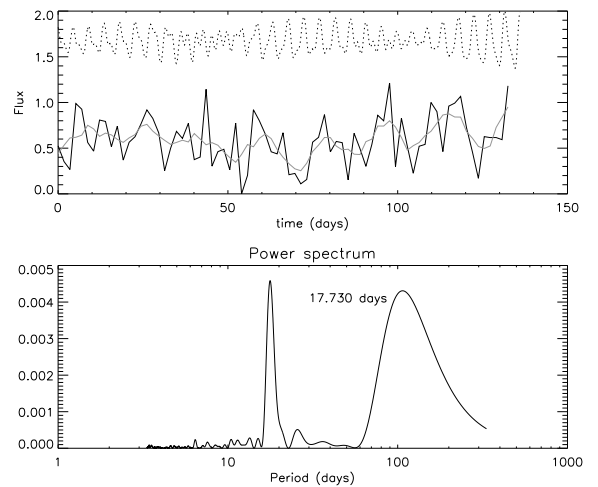}
      \caption{Top: Longitude-integrated flux deficit of all the spots in each transit. The thick line represents a running mean every 5 transits. Bottom: Power spectrum of the total flux deficit plotted in the top panel.}
         \label{lc}
\end{figure}

The time profile of the spot flux deficit integrated over all rotational longitudes  is plotted in the top panel of Figure~\ref{lc} and is independent of the adopted stellar rotation period. A running mean every 5 points is overploted as a thick line. A periodicity is clearly seen in this integrated variation. Its Lomb-Scargle periodogram is shown in the bottom panel of Figure~\ref{lc} and a prominent peak is seen at $17.7 \pm 2.3$ days. Coincidentally, this period is close to 10 times the orbital period of the planet, i.e., 17.43 days, or approximately 4 times the stellar rotation period in the occulted band, {\it i.e}, 17.92 days.

\begin{figure}
   \centering
  \includegraphics[width=8cm]{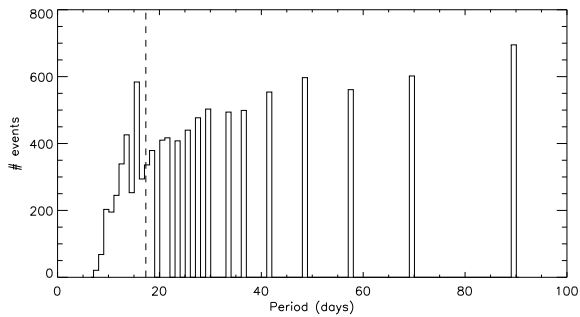}
      \caption{Histogram of the periods calculated from the  longitude-integrated total flux deficit when the order of transits is shuffled. The dashed line at P=17.33 days represents the periodicity of the observed time series.}
         \label{shuffle}
\end{figure}

To determine whether the 17.7~d periodicity is real, we performed a random permutation of the transits, and recalculated the periodogram using the IDL SCARGLE routine with a maximum period of 100 days. A total of 10,000 permutations were done and the  histogram of the resulting periods is plotted in Figure~\ref{shuffle}. As can be seen from this plot, the probability of a chance occurrence of the  periodicity at 17.7 days (marked by the dashed line) is $\sim 3$\%, therefore, we conclude that this periodicity should not be an artefact of the observational sampling, even though it is very close to 10 times the orbital period.

The periodogram of the variation of the total spotted area as derived from the modelling of the out-of-transit light curve by \citet{lanza09} is shown in their Figure~7. It shows a highly significant periodicity at $29 \pm 4$ days, whereas  the third peak in order of decreasing power corresponds to a period of 8.7 days, i.e., to the first harmonic of the periodicity of the total spotted area in the occulted band. Although its power is significantly below the level corresponding to a false-alarm probability of 1\%,  a component of the spot area modulation with that period cannot be excluded. The independent modelling by \citet{huber10} confirms the variation of the total spotted area vs. time found by \citet{lanza09}, giving support to their results. 

\section{Differential rotation} 

The analysis of the out-of-transit data yields a stellar rotation period of 4.54 days \citep{alonso08}. \cite{lanza09} obtain a slightly shorter period of 4.522 days. While the out-of-transit period is an average over the entire latitude range covered by the starspots, the period from our spot mapping  refers to a restricted band of the stellar surface, that is, only the latitudes probed by the planetary transit, $-14 \fdg 6 \pm 10^\circ$. Therefore, the period of 4.48~d found here refers to a latitude close to the equator and implies a differential rotation. This differential rotation could not have been detected if the dominant region free of spots, analogous with a solar coronal hole, had not been ignored. 

\begin{figure}
   \centering
  \includegraphics[width=8cm]{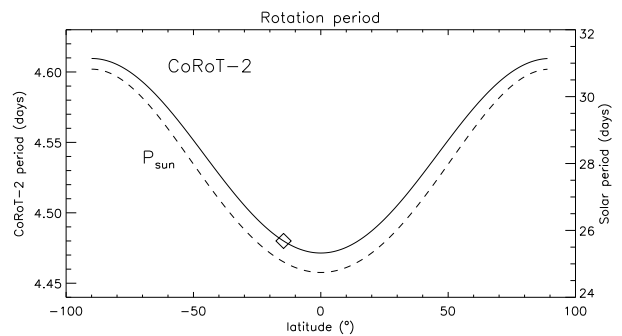}
      \caption{Differential rotation curve versus latitude for CoRoT-2 (solid line) and the Sun (dotted line).  The diamond shows the value of the rotation period of the transit latitudes.}
         \label{rotper}
   \end{figure}

The Sun rotates differentially with an angular velocity law given by Eq.~(\ref{eq:omega}). This angular velocity can be converted into a period as a function of co-latitude, $\theta$:

\begin{equation}
P_{\rm Sun} = {360^\circ  \over A - B \cos^2\theta}\\ {\rm days}, 
\label{eq:psun}
\end{equation}

\noindent where $A$ and $B$ are the same constants as in Eq.~(\ref{eq:omega}). To obtain the equivalent law for the  differential rotation on CoRoT-2, a function similar to Eq.~(\ref{eq:psun}) is used. This assumption is supported by the Doppler imaging observations discussed by \citet{barnesetal05} that show that a solar-like  differential rotation law is generally appropriate to fit  the variation of the angular velocity vs. latitude in late-type stars. In the case of CoRoT-2,  the values of $A$ and $B$ are determined by matching the period of 4.48 days where the co-latitude is $\theta = 90^\circ-\alpha = 75.4^\circ$, and 4.54 days where this function is averaged over all values of $\theta$.
These two conditions yield  $A = (80\fdg 5 \pm 0 \fdg 3)$ day$^{-1}$ and $B = (2\fdg 9 \pm 0\fdg 8)$ day$^{-1}$, thus:

\begin{equation}
P_{\rm CoRoT-2} = {360^\circ  \over 80.5 - 2.9 \cos^2(90^\circ - \alpha)}\\ {\rm days}, 
\label{eq:p2}
\end{equation}

\noindent where $\alpha$ is the latitude. The differential rotation of CoRoT-2 given by Eq.~(\ref{eq:p2}) is plotted in Figure~\ref{rotper} as a function of the  latitude. For comparison, the  differential rotation of the Sun is plotted in the figure as a dashed line (scale on the right side of the plot). The diamond marks the data point obtained from the transit spot mapping, that is, the CoRoT-2 period  of 4.48 days at $\alpha = -14 \fdg 6$.

\section{Discussion and conclusions}
\label{discussion}

A total of 77 transits of the planet around CoRoT-2 were analysed and the variations detected on the transit light curves were modelled as being due to flux deficits caused by circular spots having all the same latitude of $-14 \fdg 6$, on the stellar surface. The physical characteristics of these spots were presented in a previous paper \citep{silva-valio10}. The spot parameters were recalculated here, yielding fewer spots, however the qualitative spots characteristics have not changed. Here we concentrate on the temporal evolution of the spots.

Before the spot temporal behaviour can be estimated, it is necessary to know the rotation period of the star. Out-of-transit data yield periods of 4.54 \citep{alonso08} or 4.522 \citep{lanza09} days. The agreement of the spots longitude and flux deficit obtained from out-of-transit and transit mapping is remarkable (see Figure~\ref{lanza}), and  provides an independent confirmation of the out-of-transit maximum entropy mapping method.

Here, the goal was to obtain the rotation period within the transit latitudes from the spot longitude map.
At first sight, it appeared that the surface rotation period of the transit latitudes was equal to the out-of-transit value, i.e., 4.54 days,  implying that the star displays no differential rotation. However, if one interprets the lack of spots, or active regions, within the longitude range between $0^{\circ}$  and 40$^\circ$ as the photospheric counterpart of a coronal hole extending to the equatorial regions (in analogy with the solar case), then it is no surprise that this region presents a rotation period similar to the mean value of 4.54 days.

In the Sun, extended unipolar magnetic regions are  often found in the photosphere below coronal holes and share their rotational properties with them. Specifically, the properties of solar coronal holes have been investigated by, e.g., \citet{shelke85}, \citet{insley95}, \citet{bumba95}, \citet{ikhsanov99}.
There are basically two classes of coronal holes, those that are confined to the polar regions and dominate during periods of minimum activity, and those that appear at mid-latitudes or even in the equatorial region, and predominate during solar maximum. The polar coronal holes may also present equator-ward extensions \citep{insley95}.
\citet{insley95} studied the differential rotation of coronal holes, and found that they rotate more rigidly than the photosphere. The mid-latitude coronal holes have an equatorial rotation period of 27.1 days, whereas the rotation of the equator ward extensions of polar coronal holes is even more rigid, showing almost no differential rotation. The rotation period at the equator of this class of coronal holes is 28.3 days \citep{insley95}. The equatorial rotation period of the mid-latitude coronal holes   is very close to the average rotation period of the solar photosphere. 

Ignoring the longitude range free of spots in CoRoT-2,  we searched for vertical strings of starspots in the maps of Figure~\ref{dif} while also inspecting the auto-correlation function of the spot time integrated flux deficit vs. longitude. According to this criterion, the best rotation period was found to be 4.48 days that minimized the drift in longitude of the individual spots and was  taken as the rotation period of the stellar surface within the $-14 \fdg 6 \pm 10^{\circ}$ latitude band. Using this rotation period for the star, the stellar differential rotation profile, shown in Figure~\ref{rotper}, was defined. This profile shows that close to the equator the star rotates faster, with a period of $4.47\pm 0.01$~d, while the longest period of $4.61 \pm 0.03$ days is found at the poles, as  obtained from Eq.~(\ref{eq:p2}). 

\cite{lanza09} found a lower limit to the amplitude of the relative differential rotation of $\sim 0.7 - 1$ \% for this star, significantly smaller than the Sun. The relative differential rotation of the Sun is 22\%. Here we obtain $(P_{\rm pole} - P_{\rm eq})/{\bar P} =$ 3\%, where $\bar P$ is the mean rotation period.  In CoRoT-2 the out-of-transit light modulation is dominated by the lack of spots within the coronal hole close to the equator,  yielding almost no differential rotation.

From a spot model with three spots only, \cite{frohlich09} find a mean differential rotation of 0.11 rad/d and a maximum differential rotation of 0.129 rad/d. Using the values obtained here, we find $\Delta \Omega = \Omega_{\rm eq} - \Omega_{\rm pole}$ = 0.042 rad/d. This is an intermediate value between that found by \cite{lanza09} and \cite{frohlich09}. Considering the results of \citet{barnesetal05}, the differential rotation of  CoRoT-2 is approximately half the pole-equator angular velocity difference generally found in late-type stars of similar effective temperature and rotation period, suggesting that its spots are concentrated in an equatorial belt, as in the Sun. 

The  temporal evolution of the spot surface coverage within  $20^\circ$ longitude bins can be estimated. The average value of the spot flux temporal variation is estimated as $31 \pm 15$ days.
\cite{lanza09}, from out-of-transit data, find that the  total spotted area of CoRoT-2 oscillates with a period of  $29 \pm 4$ days, close to the mean spot flux deficit periodicity estimated here.

Another interesting feature is the temporal evolution of the average flux deficit, i.e., integrated over longitude in the occulted band. It shows a periodicity of $17.7 \pm 2.3$ days with a false-alarm probability of $\sim 3$ \%. This is very close to 10 times the orbital period of the planet or approximately 4 times the stellar rotation period found within the transit band. 

These results may suggest some kind of magnetic interaction between the star and the planet. 
 Circumstantial evidence for  star-planet interaction  in the magnetic activity of the photosphere has been reviewed by \citet{lanza08,lanza09m} and \citet{shkolniketal09}. A major problem is the lack of a common and well-defined phenomenology in all the stars suspected to show such an interaction. Therefore, our suggestion should be regarded only as a tentative interpretation of the periodicities found for the total spotted area. 

An alternative interpretation  may be that  proposed by \citet{lou00} to account for the short-term cycles observed in the Sun close to  some of the maxima of the 11-yr cycles, the so-called Rieger cycles \citep{riegeretal84}. For instance, \citet{oliveretal98} detected oscillations with a period of $\sim 158$ days in the total sunspot area close to the maximum of solar cycle 19.  \citet{lou00} interprets such oscillations as signatures of hydromagnetic Rossby-type waves trapped in the subphotospheric layers of the star. In the case of the Sun, the main periodicity corresponds to $\sim 6$ equatorial rotation periods.  Since the rotation period of  CoRoT-2 is only  4.5 days, this would correspond to a wave period of $\sim 27$ days, close to the periodicity of $28.9$ days found by \citet{lanza09} from the modelling of the out-of-transit light modulation. 
Moreover, other periodicities, ranging from 2 to 6 rotation periods, have  been reported from the analysis of different solar data sets \citep{lou00}. In the case of CoRoT-2, we found in Sect.~\ref{time_evolution} a periodicity of 17.7 days in the  total spotted area of the occulted band, very close to 4 stellar rotation periods at that latitude, which would tentatively correspond to the 102-day periodicity reported by \citet{lou00} for the Sun. Its first harmonic, corresponding to 2 rotation periods, could also be present in the Sun as well as in the variation of the total spotted area of CoRoT-2  derived from its out-of-transit light modulation (cf. Sect.~\ref{time_evolution}).

\begin{acknowledgements}
We would like to thank everyone involved in the planning and operation of the CoRoT satellite which made these observations possible. A.S.V. acknowledges partial financial support from the Brazilian agency FAPESP (grant number 2006/50654-3).
\end{acknowledgements}


\begin{thebibliography}{}

\bibitem[Alonso et al.(2008)]{alonso08} Alonso, R., Auvergne, A., Baglin, A. et al. 2008, A\&A, 482, L21

\bibitem[Auvergne et al.(2009)]{auvergneetal09} 
Auvergne, M., et al.\ 2009, \aap, 506, 411 

\bibitem[Barnes et al.(2005)]{barnesetal05} 
Barnes, J.~R., Cameron, A.~C., Donati, J.-F., et al. 2005, \mnras, 357, L1 

\bibitem[Berdyugina(2005)]{berdyugina05}
Berdyugina, S. V. 2005, Living Rev. Solar Phys., 2, 8 
(http://www.livingreviews.org/lrsp-2005-8)

\bibitem[Bouchy et al.(2008)]{bouchy08} 
Bouchy, F., Queloz, D., Deleuil, M., et al. 2008, A\&A, 482, L25

\bibitem[Bumba et al.(1995)]{bumba95}
Bumba, V., Klvana, M., \& S\'ykora, J. 1995, \aap, 298, 923

\bibitem[Carrington(1863)]{carr1863} Carrington, R. C. 1863, Observations of the Spots on the Sun (London: Williams \& Norgate).

\bibitem[Charbonneau(2000)]{charbonneau00}
Charbonneau, D., Brown, T. M., Latham, D. W., \& Mayor, M. 2000, \apj, 529, L45

\bibitem[Charbonneau(1995)]{charbonneau95}
Charbonneau, P. 1995, \apjs, 101, 309

\bibitem[Fr\"ohlich et al.(2009)]{frohlich09}
Fr\"ohlich, H.-E., K\"uker, M., Hatzes, A. P., \& Strassmeier, K. G. 2009, A\&A, 506, 263

\bibitem[Gillon et al.(2010)]{gillonetal10} 
Gillon, M., et al.\ 2010, \aap, 511, A3 

\bibitem[Henry et al.(2000)]{henry00}
Henry, G. W., Marcy, G. W.,  Butler, R. P., \& Vogt, S. S. 2000, \apj, 529, L41

\bibitem[Huber et al.(2009)]{huberetal09}
 Huber, K.~F., Czesla, S., Wolter, U., \& Schmitt, J.~H.~M.~M.\ 2009, \aap, 508, 901 

\bibitem[Huber et al.(2010)]{huber10}
Huber, K. F., Czesla, S., Wolter, U., \& Schmitt, J. H. M. M. 2010, \aap, 514, A39 

\bibitem[Ikhsanov \& Ivanov(1999)]{ikhsanov99}
Ikhsanov, R. N. \& Ivanov, V. G. 1999, Solar Phys., 188,245

\bibitem[Insley et al.(1995)]{insley95}
Insley, J. E., Moore, V., \& Harrison, R. A. 1995, Solar Phys., 160, 1

\bibitem[Kron(1947)]{kron47}
Kron, G. E. 1947, \pasp, 59, 261

\bibitem[Lampton et al.(1976)]{lampton76}
Lampton, M., Margon, B., \& Bowyer, S.\ 1976, \apj, 208, 177

\bibitem[Lanza(2008)]{lanza08} 
Lanza, A.~F.\ 2008, \aap, 487, 1163 

\bibitem[Lanza(2009)]{lanza09m} 
Lanza, A.~F.\ 2009, \aap, 505, 339 

\bibitem[Lanza et al.(2009)]{lanza09} 
Lanza, A. F., Pagano, I., Leto, G., et al. 2009, A\&A, 493, 193

\bibitem[Lou(2000)]{lou00} 
Lou, Y.-Q.\ 2000, \apj, 540, 1102 

\bibitem[Oliver et al.(1998)]{oliveretal98} 
Oliver, R., Ballester, J.~L., \& Baudin, F.\ 1998, \nat, 394, 552 

\bibitem[Press et al.(1992)]{press92} 
Press, W. J., Teukolsky, S. A., Vetterling, W. T. \& Flannery, B. P. 1992, Numerical recipes in FORTRAN, The art of scientific computing (Cambridge: University Press), 2nd. Edition

\bibitem[Rieger et al.(1984)]{riegeretal84} 
Rieger, E., Kanbach, G., Reppin, C., et al. 1984, \nat, 312, 623 

\bibitem[Shelke \& Pande(1985)]{shelke85}
Shelke, R. N. \& Pande, M. C. 1985, Solar Phys., 95,193

\bibitem[Shkolnik et al.(2009)]{shkolniketal09} 
Shkolnik, E., et al.\ 2009, American Institute of Physics Conference Series, 1094, 275 

\bibitem[Silva(2003)]{silva03} 
Silva, A. V. R. 2003, \apj, 585, L147

\bibitem[Silva-Valio(2008)]{silva-valio08} 
Silva-Valio, A. 2008, \apj, 683, L179 

\bibitem[Silva-Valio et al.(2010)]{silva-valio10} 
Silva-Valio, A., Lanza, A. F., Alonso, R. \& Barge, P. 2010, \aap, 510, A25 

\bibitem[Strassmeier(2009)]{strassmeier09} 
Strassmeier, K. G.\ 2009, \aapr, 17, 251 

\bibitem[Thomas \& Weiss(2008)]{thomas08}
Thomas, J. H. \& Weiss, N. O. 2008, Sunspots and Starspots (Cambridge: University Press), p. 189

\bibitem[Wolter et al.(2009)]{wolteretal09} 
Wolter, U., Schmitt, J.~H.~M.~M., Huber, K.~F., et al. 2009, \aap, 504, 561 



\end{thebibliography}
\end{document}